\documentclass{aa}
\usepackage{graphicx}
\usepackage{txfonts}
\usepackage{hyperref}
\usepackage{multirow}
\usepackage[capitalise]{cleveref}
\usepackage[per-mode=symbol]{siunitx}
\usepackage{graphicx}
\usepackage{textcomp}
\usepackage{color}
\bibpunct{(}{)}{;}{a}{}{,}

\DeclareSIUnit{\astronomicalunit}{au}

\definecolor{RED}{RGB}{255,0,0}

\newcommand{\md}{\text{d}}
\newcommand{\nut}{{\tilde \nu}}

\definecolor{RED}{RGB}{255,0,0}

\begin{document}

\title{Treatment of overlapping gaseous absorption with the correlated-$k$ method in hot~Jupiter and brown~dwarf atmosphere models}

\titlerunning{Treatment of overlapping absorption in hot~Jupiter and brown~dwarf atmosphere models}

\author{%
David~S.~Amundsen\inst{1,2,3}\and%
Pascal~Tremblin\inst{1,4}\and%
James~Manners\inst{1,5}\and%
Isabelle~Baraffe\inst{1,6}\and%
Nathan~J.~Mayne\inst{1}%
}
\authorrunning{Amundsen  et al.}

\institute{%
Astrophysics Group, University of Exeter, Exeter, EX4 4QL, United Kingdom
\and
Department of Applied Physics and Applied Mathematics, Columbia University, New York, NY 10025, USA\\
\email{d.s.amundsen@columbia.edu}
\and
NASA Goddard Institute for Space Studies, New York, NY 10025, USA
\and
Maison de la Simulation, CEA-CNRS-INRIA-UPS-UVSQ, USR 3441, Centre d'\'{e}tude de Saclay, F-91191 Gif-Sur-Yvette, France
\and
Met Office, Exeter, EX1 3PB, United Kingdom
\and
Univ Lyon, ENS de Lyon, Univ Lyon 1, CNRS, CRAL, UMR5574, F-69007, Lyon, France
}

\date{}

 

\abstract{
The correlated-$k$ method is frequently used to speed up radiation calculations in both one-dimensional and three-dimensional atmosphere models. An inherent difficulty with this method is how to treat overlapping absorption, i.e. absorption by more than one gas in a given spectral region. We have evaluated the applicability of three different methods in hot Jupiter and brown dwarf atmosphere models, all of which have been previously applied within models in the literature: (i) Random overlap, both with and without resorting and rebinning, (ii) equivalent extinction and (iii) pre-mixing of opacities, where (i) and (ii) combine $k$-coefficients for different gases to obtain $k$-coefficients for a mixture of gases, while (iii) calculates $k$-coefficients for a given mixture from the corresponding mixed line-by-line opacities. We find that the random overlap method is the most accurate and flexible of these treatments, and is fast enough to be used in one-dimensional models with resorting and rebinning. In three-dimensional models such as GCMs it is too slow, however, and equivalent extinction can provide a speed-up of at least a factor of three with only a minor loss of accuracy while at the same time retaining the flexibility gained by combining $k$-coefficients computed for each gas individually. Pre-mixed opacities are significantly less flexible, and we also find that particular care must be taken when using this method in order to to adequately resolve steep variations in composition at important chemical equilibrium boundaries. We use the random overlap method with resorting and rebinning in our one-dimensional atmosphere model and equivalent extinction in our GCM, which allows us to e.g. consistently treat the feedback of non-equilibrium chemistry on the total opacity and therefore the calculated $P$--$T$ profiles in our models.
}


\maketitle

\section{Introduction}

Rapid calculation of wavelength-integrated fluxes and heating rates are needed in most planetary and brown dwarf atmosphere models. As line-by-line approaches are too computationally expensive for practical use, the correlated-$k$ method~\citep{Goody1989,Lacis1991,Thomas2002} has been applied in retrieval models~\citep[e.g.][]{Irwin2008}, one-dimensional (1D) atmosphere models~\citep[e.g.][]{Marley1996,Burrows1997,Fortney2005}, and three-dimensional (3D) global circulation models (GCMs)~\citep[e.g.][]{Showman2009,Kataria2013,Amundsen2016a}. With the correlated-$k$ method the spectrum is divided into bands and, in each band, the opacity probability distribution is derived and described by a small number (usually $8$ to $16$) of $k$-coefficients and corresponding weights. Pseudo-monochromatic calculations are performed using these $k$-coefficients, decreasing the required computation time by several orders of magnitude compared to line-by-line calculations.

The treatment of overlapping gaseous absorption, i.e. absorption by more than one gas in a single spectral interval, with the correlated-$k$ method is a difficult issue. Bands can be chosen such that absorption is dominated by a single gas in each band, however, this choice will be imperfect both because the relative strength of absorbers may change with temperature and pressure and due to spectral regions with significant overlap of the absorption of different gases. It is therefore necessary to take into account absorption by more than one gas in the same spectral interval. Several different schemes for deriving $k$-coefficients for gas mixtures have been developed for the Earth atmosphere~\citep[see e.g.][]{Goody1989,Lacis1991,Fu1992,Edwards1996b,Buchwitz2000,Yang2000,Li2005,Shi2009,Hogan2010,Sun2011}, each with advantages and drawbacks. The goal of this paper is not to review each of these, but to evaluate the accuracy and flexibility of three schemes that have previously been applied in hot Jupiter and brown dwarf atmosphere models:
\begin{enumerate}
\item
\textit{Pre-mixed $k$-coefficients}~\citep[PM,][]{Goody1989}: $k$-coefficients for the mixture are computed directly from the total line-by-line gas opacity.
\item
\textit{The random overlap method}, both without (RO) and with (RORR) resorting and rebinning~\citep{Lacis1991}: $k$-coefficients are computed for each gas and combined assuming their absorption cross-sections are uncorrelated.
\item
\textit{Equivalent extinction}~\citep[EE,][]{Edwards1996b}: $k$-coefficients are computed for each gas and combined using an ``equivalent grey absorption'' for all minor absorbers and all $k$-coefficients for the major absorber in each band.
\end{enumerate}

Pre-mixed $k$-coefficients have been employed in solar system planet, exoplanet and brown dwarf atmosphere models~\citep[see e.g.][]{Marley1996,Burrows1997,Fortney2005,Showman2009,Wordsworth2013}. This method avoids problems related to combining $k$-coefficients for different gases, but is inflexible as mixing must be assumed before $k$-coefficients are computed. Alternatively, gas mixing ratios can be added as dimensions to the look-up table of $k$-coefficients, however, this leads to a very large number of dimensions in the table.

The random overlap method has been applied in retrieval models~\citep{Irwin2008} and 1D brown dwarf atmosphere models~\citep{Tremblin2015,Tremblin2016}, and assumes that the absorption cross-sections of different gases are uncorrelated. The total number of $k$-coefficients in a band scales as the product of the number of $k$-coefficients for each overlapping gas, causing this method to become computationally expensive, but resorting and rebinning the resulting $k$-coefficients can be used to circumvent this issue~\citep{Lacis1991}. We have recently applied equivalent extinction in our GCM to study hot Jupiters~\citep{Amundsen2016a}. Like the random overlap method this method is more flexible than using pre-mixed $k$-coefficients, but requires knowledge of which absorbers should be treated as the major and minor sources of opacity in each band.

It is clearly beneficial in terms of model flexibility to compute $k$-coefficients individually for each gas and combine them on-the-fly in models using the current local mixing ratios. As all wavelength information is lost when the $k$-coefficients are computed it is impossible to do this perfectly without loss of accuracy, and requires an assumption about the absorption of the different gases. The random overlap method assumes that the lines of different gases are randomly overlapping (or equivalently that the absorption cross-sections are uncorrelated), while equivalent extinction assumes minor absorbers can be treated as grey. It is essential to verify the accuracy of these assumptions by comparing to line-by-line calculations.

In this paper we compare pre-mixing, random overlap and equivalent extinction in terms of computational efficiency and evaluate their accuracy by comparing to results from line-by-line calculations. In \cref{sec:corr-k} we give a brief overview of the correlated-$k$ method and \cref{sec:overlap} describes the above overlap schemes in more detail. In \cref{sec:application} we apply them in hot Jupiter atmosphere models, compare them and evaluate their computational efficiency, by using our 1D radiative-convective equilibrium atmosphere code \texttt{ATMO}~\citep{Tremblin2015,Tremblin2016} and our GCM radiation scheme SOCRATES\footnote{\url{https://code.metoffice.gov.uk/trac/socrates}}~\citep{Edwards1996a,Edwards1996b,Amundsen2014}. We give our concluding remarks in \cref{sec:conclusions}.

\section{The correlated-$k$ method} \label{sec:corr-k}

As treating the wavelength-dependence of gaseous absorption explicitly is too computationally expensive to be performed in many atmosphere models, the correlated-$k$ method is frequently used. It considers the probability distribution of the opacity in the spectral bands and assumes that the mapping between spectral regions and the probability distribution is vertically correlated. Originally developed for the Earth atmosphere~\citep{Lacis1991}, it has since been adopted in both one-dimensional~\citep{Marley1996,Burrows1997,Marley2014,Tremblin2015} and global circulation models~\citep{Showman2009,Kataria2013,Amundsen2016a} of hot Jupiter and brown dwarf atmospheres. We do not discuss the correlated-$k$ method in detail here, but refer to e.g. \citet{Lacis1991}, \citet{Goody1989} and \citet{Thomas2002} for in-depth discussions. Note that we have previously verified the applicability of the correlated-$k$ method in hot Jupiter and brown dwarf atmosphere models~\citep{Amundsen2014}.

In the correlated-$k$ method the opacity spectrum is divided into bands $b$. In each band $k$-coefficients $k_l^b$ and corresponding weights $w_l^b$ are computed from the probability distribution of the opacity, with $l \in [1,n_k^b]$ where $n_k^b$ is the number of $k$ coefficients within band $b$. The transmission through a homogeneous slab is given by
\begin{align}
\mathcal T(u) &= \int_{\nut_1}^{\nut_2} \md \nut \, w(\nut) e^{-k(\nut) u}
= \int_0^1 \md g \, e^{-k(g) u}
\label{eq:transmission_def} \\
&\approx \sum_{l=1}^{n_k^b} w_l^b e^{-k_l^b u},
\label{eq:transmission}
\end{align}
where $\nut$ is the wavenumber, $\nut_1$ and $\nut_2$ are wavenumber limits of band $b$, $w(\nut)$ is a weighting function, and $k(\nut)$ and $u$ are the opacity and column density of the gas, respectively. $g(k)$ is the cumulative opacity probability distribution, where $g(k)$ is the probability of having an opacity $\leq k$ within the band.

Pseudo-monochromatic fluxes $F_l^b$ are computed for each $k_l^b$-coefficient, with the integrated flux in band $b$ given by
\begin{equation}
F^b = \sum_{l=1}^{n_k^b} w_l^b F_l^b,
\label{eq:flux_band}
\end{equation}
and the total spectral integrated flux given by
\begin{equation}
F = \sum_{b=1}^{n_b} F^b,
\label{eq:flux}
\end{equation}
where $n_b$ is the number of bands.

The $k_l^b$-coefficients are the $k$-coefficients for the gas mixture, i.e. taking into account all absorbers present. Spectral bands can be chosen such that absorption is dominated by only one gas, the major absorber, in each band. Other gases may still contribute significantly to absorption, however, which causes the need to treat overlapping absorption. In addition, in some spectral regions the major and minor absorbers may change depending on the gas mixing ratios. Consequently, there is a need to compute $k$-coefficients for a gas mixture.

\section{Treatments of gaseous overlap} \label{sec:overlap}

In this section we briefly discuss three different methods for treating overlapping gaseous absorption previously used in hot Jupiter and brown dwarf atmosphere models in the literature. For convenience we adopt a set of acronyms for these overlap methods, which we summarise in \cref{tbl:acronyms}.

\begin{table}
\centering
\caption{Methods for treating overlapping gaseous absorption considered here and our adopted acronyms.}
{\small
\begin{tabular}{l|l}
Overlap method & Acronym \\ \hline
Random overlap & RO \\
Random overlap with resorting and rebinning & RORR \\
(Adaptive) Equivalent extinction & (A)EE \\
Pre-mixing & PM
\end{tabular}
}
\label{tbl:acronyms}
\end{table}

\subsection{Pre-mixed}

The total absorption coefficient can be calculated by summing line-by-line absorption coefficients for all absorbing species weighted by their relative abundances:
\begin{equation}
k^\text{tot}(\nut,P,T) = \sum_{i=1}^{N_\text{s}} k_i(\nut,P,T) \zeta_i(P,T),
\end{equation}
where the sum is over all $N_\text{s}$ species, and $k_i(\nut,P,T)$ and $\zeta_i(P,T)$ are the absorption coefficient and mixing ratio of gas $i$ at pressure $P$ and temperature $T$, respectively. The total absorption coefficient at a given $(P,T)$ is then given by $k^\text{tot} \rho$, where $\rho$ is the total gas density. $k^\text{tot}$ can be used to compute and tabulate $k$-coefficients for the gas mixture as a function of temperature and pressure. This approach has several advantages: it is fast, requiring only one set of $k$-coefficients for each temperature and pressure, and it is simple to implement. This technique has been used in 1D atmosphere models \citep[e.g.][]{Marley2014} and the SPARC/MITgcm \citep{Showman2009}. It is not particularly flexible, however, as the local mixing ratios $\zeta_i(P,T)$ must be determined before the time consuming calculation of $k$-coefficients. A potential solution would be to add gas mixing ratios as dimensions to the look-up table of $k$-coefficients, but the increased size of such a table is prohibitive for application in atmosphere models with many absorbing gases.

\subsection{The random overlap method}

The second method we discuss is the random overlap method~\citep{Lacis1991}. $k$-coefficients are computed individually for each gas, and $k$-coefficients for different gases are combined assuming that the absorption coefficient of one gas $x$, is uncorrelated to that of a second gas $y$, i.e. that their lines are randomly overlapping. The total transmission of the gas mixture over some column density $(u_x,u_y)$ is then given by a simple scalar product,
\begin{equation}
\mathcal T(u_x, u_y) = \mathcal T_x(u_x) \times \mathcal T_y(u_y) .
\label{eq:T_ab}
\end{equation}
We show in \cref{sec:convolution} that this is equivalent to convolving the opacity probability distributions of the different gases. The assumption that the absorption coefficients are uncorrelated will depend on the adopted bands and its applicability should be verified by comparing to line-by-line calculations. We perform such a comparison in \cref{sec:random_overlap}.

\subsubsection{Without resorting and rebinning}

\begin{figure*}
\centering
\includegraphics[width=0.8\textwidth]{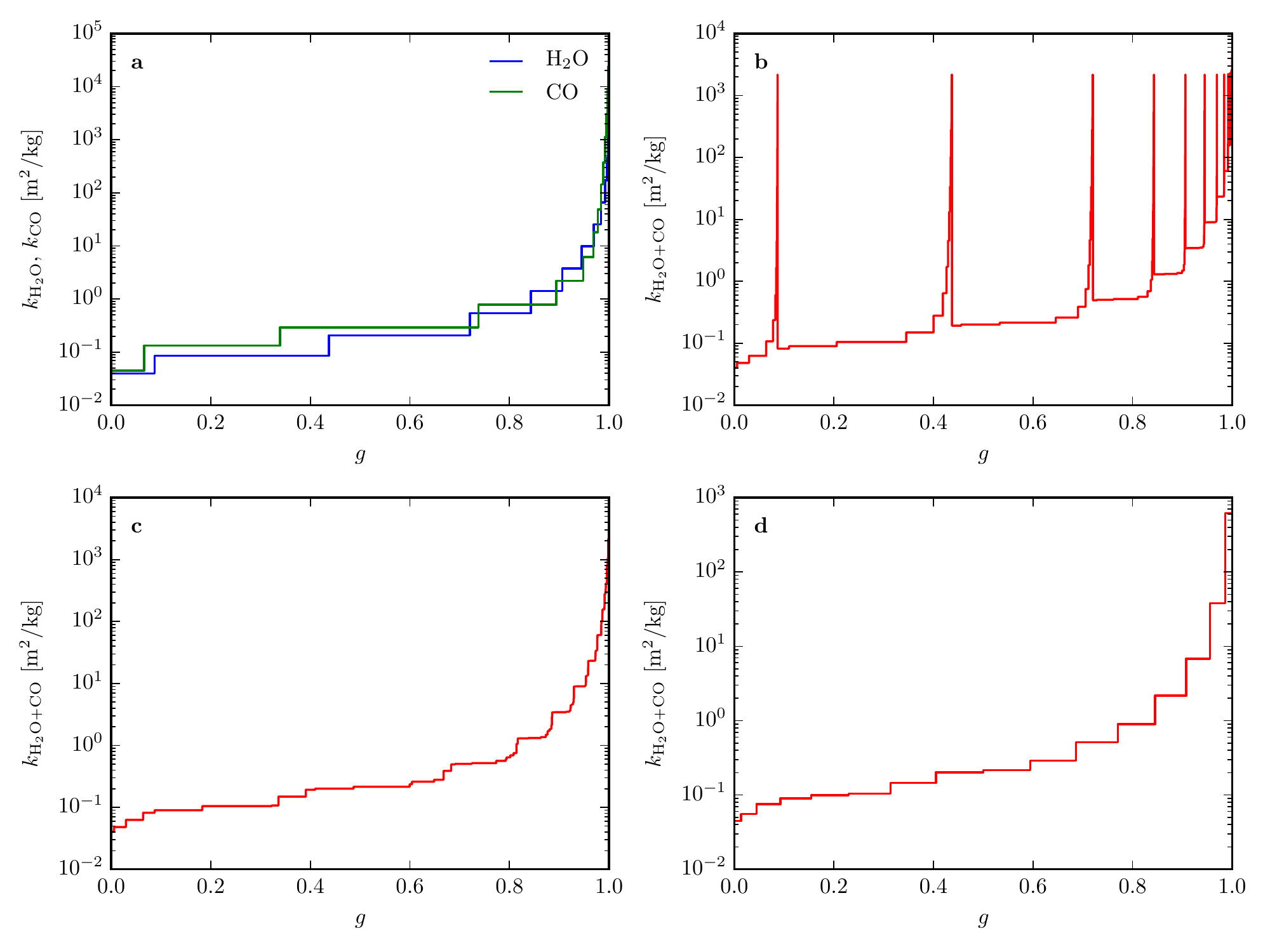}
\caption{Schematic illustration of the random overlap method. The original $k$-distributions of H$_2$O and CO are shown in panel~a, with the combined $k$-coefficients shown in panel~b calculated using \cref{eq:k_ro,eq:w_ro} assuming $\zeta_\text{H$_2$O} = 0.9$ and $\zeta_\text{CO} = 0.1$. Without resorting and rebinning (RO) the $k_{xy,lm}$-coefficients with corresponding weights $w_{xy,lm}$ are used directly. With resorting and rebinning (RORR) the combined $k$-coefficients in panel~b are resorted, as shown in panel~c, and then rebinned down to a smaller number of $k$-coefficients, as shown in panel~d.}
\label{fig:rorr_schematic}
\end{figure*}

\Cref{eq:T_ab} can be rewritten in terms of the $k$-coefficients for the individual gases $x$ and $y$. The transmission through one layer is, using \cref{eq:transmission_def,eq:transmission,eq:T_ab},
\begin{align}
\mathcal T(u_x, u_y) &= \int_{\nut_1}^{\nut_2} \md \nut \, w_x(\nut) e^{-k_x(\nut) u_x} \times \int_{\nut_1}^{\nut_2} \md \nut' \, w_y(\nut) e^{- k_y(\nut') u_y} \\
&= \sum_{l=1}^{n_{k,x}} \sum_{m=1}^{n_{k,y}} w_{x,l} w_{y,m} e^{-k_{x,l} u_x -k_{y,m} u_y}.
\end{align}
Defining $u_{xy} = u_x + u_y$, we can write the above transmission as
\begin{equation}
\mathcal T(u_x, u_y) = \sum_{l=1}^{n_{k,x}} \sum_{m=1}^{n_{k,y}} w_{xy,lm} e^{-k_{xy,lm} u_{xy}},
\end{equation}
where
\begin{align}
k_{xy,lm} &= \frac{k_{x,l} u_x + k_{y,m} u_y}{u_x + u_y}
= \frac{k_{x,l} \zeta_x u + k_{y,m} \zeta_y u}{\zeta_x u + \zeta_y u} \\
&= \frac{k_{x,l} \zeta_x + k_{y,m} \zeta_y}{\zeta_x + \zeta_y},
\label{eq:k_ro}
\end{align}
and
\begin{equation}
w_{xy,lm} = w_{x,l} w_{y,m} .
\label{eq:w_ro}
\end{equation}
For illustration we show schematics of the $k$-coefficients H$_2$O and CO in \cref{fig:rorr_schematic}\textbf{a}, with the combined $k$-coefficients in \cref{fig:rorr_schematic}\textbf{b} calculated using \cref{eq:k_ro} assuming $\zeta_\text{H$_2$O} = 0.9$ and $\zeta_\text{CO} = 0.1$.

Running $n_{k,x} n_{k,y}$ pseudo-monochromatic calculations using the $k_{xy,lm}$-coefficients the total flux can be calculated as usual using \cref{eq:flux_band,eq:flux} with the weights $w_{xy,lm}$. This procedure can be replicated for an arbitrary number of gases, however, the computation time increases by a factor of $n_{k}$ for each gas added. This method therefore quickly becomes too computationally expensive for practical use.

\subsubsection{With resorting and rebinning} \label{sec:rorr}

\citet{Lacis1991} suggested that ranking and reblocking, i.e. resorting the $k_{xy,lm}$-coefficients and rebinning them to obtain a smaller number of $k$-coefficients $k_{xy,l}^\text{bin}$, would circumvent the scaling issue. The procedure is illustrated in panels c and d of \cref{fig:rorr_schematic}. First the $k$-coefficients of two gases are combined using \cref{eq:k_ro,eq:w_ro}, as shown in \cref{fig:rorr_schematic}b. These $n_{k,x} n_{k,y}$ $k$-coefficients are sorted in increasing order, with the weights sorted using the same mapping, as shown in \cref{fig:rorr_schematic}c\footnote{We have used quicksort, shellsort and heapsort, all available as standard library routines~\citep[e.g.][]{press2007numerical}, and found that quicksort is generally the fastest. We adopt quicksort in the current work.}. The sorted $k_{xy,lm}$-coefficients are then binned down to $n_k^\text{bin}$ $k_{xy,l}^\text{bin}$-coefficients, which we show in \cref{fig:rorr_schematic}d.

The weights used in the rebinning, $w_{xy,l}^\text{bin}$, must be the same for all layers, and should ideally be spaced equally in $\log k_{xy,lm}$. As $k_{xy,lm}$ will vary vertically, however, equal spacing in $\log k_{xy,lm}$ in one layer will not correspond to equal spacing in $\log k_{xy,lm}$ at a different level. Consequently a compromise must be made, with one possible solution being an equal spacing in $\log k_{xy,lm}$ defined where the optical depth in each band reaches unity. As the above procedure of resorting and rebinning is repeated to include more than two gases, however, $\log k_{xy,lm}$ will change meaning the ideal spacing to change. In addition, particular care must be taken to treat cases where $k$-coefficients of gases are zero.

For simplicity and ease of implementation we use a much simpler approach for determining the weights $w_{xy,l}^\text{bin}$: In SOCRATES we use weights given by a Gauss-Legendre quadrature, while in \texttt{ATMO} we use uniform weights, both of which can have an arbitrary number of rebinned $k$-coefficients $n_k^\text{bin}$. The rebinned coefficients $k_{xy,l}^\text{bin}$ are found by computing a weighted average of all $k_{xy,lm}$-terms belonging to each bin $w_{xy,l}^\text{bin}$, where $w_{xy,lm}$ are used as weights. If a $k_{xy,lm}$-term extends over more than a single $w_{xy,l}^\text{bin}$ bin, it is split over neighbouring bins such that the weights $w_{xy,lm}$ sum up to exactly $w_{xy,l}^\text{bin}$ in each bin. We use a linear average, but note that the accuracy can be improved somewhat by averaging in $\log k_{xy,lm}$. This causes a significant increase in computation time, however, due to the need to compute the logarithm of many $k$-terms, and would also require particular care to treat cases where $k_{xy,lm}$-terms are zero.

After this resorting and rebinning, the process is repeated, adding one gas at a time, until all gases have been added. The final binned $k$-coefficients are used to compute the fluxes and heating rates for the atmosphere. This approach is consequently much more flexible than pre-mixing gases as gas abundances can be set at run-time.

\subsection{Equivalent extinction}

The last method of treating gaseous overlap that we consider is equivalent extinction~\citep{Edwards1996b}. It utilizes the fact that in most bands there is a primary (major) absorber, and includes additional absorbers through a grey ``equivalent extinction''. In each layer and band an equivalent extinction $\bar k$ is calculated for each minor gas, which for the thermal component is defined as
\begin{equation}
\bar k_x = \frac{\sum_{l=1}^{n_{k,x}} w_{x,l} k_{x,l} F_{\text{v},l}}{\sum_{l=1}^{n_{k,x}} w_{x,l} F_{\text{v},l}},
\label{eq:kBarApprox_net}
\end{equation}
where $k_{x,l}$ are the $k$-coefficients of the minor gas in the layer with corresponding weights $w_{x,l}$, and $F_{\text{v},l}$ is the thermal flux in the layer including only absorption by $k$-term $l$ of the gas. Pseudo-monochromatic calculations are performed for all $n_{k}$ $k$-coefficients of the major gas in each band, with all other absorbers included by using the equivalent grey absorption $\bar k_x$. This effectively reduces the number of pseudo-monochromatic calculations required to one per $k$-coefficient per gas.

The direct component of the stellar flux is readily included by calculating the transmission for each gas separately and then taking the product since, assuming random overlap, direct transmissions are multiplicative (see \cref{eq:T_ab}). For the diffuse stellar beam, which will be non-zero if scattering is included, the equivalent extinction is defined by
\begin{equation}
\bar k_x = \frac{\sum_{l=1}^{n_{k,x}} w_{x,l} k_{x,l} F_{\text{s}*,l}}{\sum_{l=1}^{n_{k,x}} w_{x,l} F_{\text{s}*,l}},
\label{eq:kBarApprox_sw}
\end{equation}
$F_{\text{s}*,l}$ is the direct flux at the lower boundary including only $k$-term $l$ of the gas. The use of $F_{\text{s}*,l}$ means that equivalent extinction in the current formulation is less suited for use in hot Jupiter atmosphere models as the direct stellar flux at the bottom boundary may be zero. In this case we use the smallest $k$-coefficient for the minor gas as $\bar k_x$. In this work, however, as we only consider Rayleigh scattering, the main stellar radiation is contained in the direct beam, making this a minor issue.

\begin{figure*}
\centering
\includegraphics[width=0.8\textwidth]{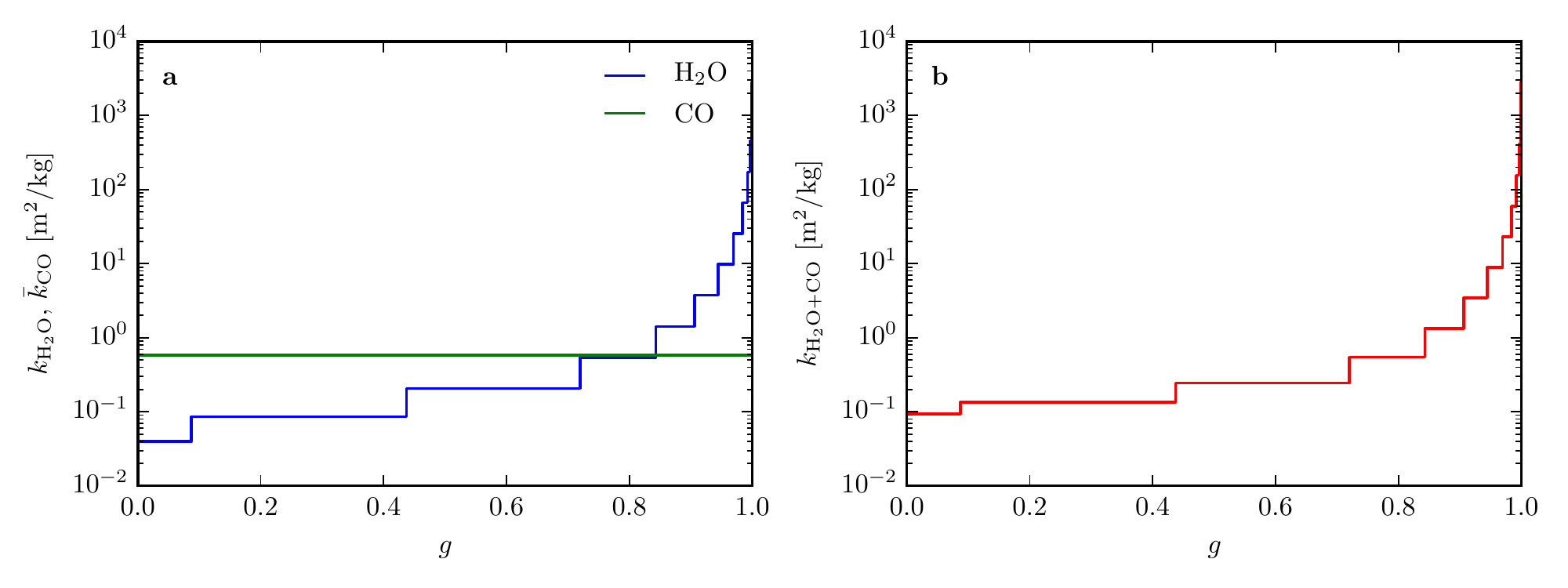}
\caption{Schematic illustration of equivalent extinction (EE). An equivalent grey absorption is calculated for each minor gas (CO in our case), as shown in panel~a, using \cref{eq:kBarApprox_net} or \cref{eq:kBarApprox_sw}. This grey absorption is added onto the $k$-coefficients of the major gas (H$_2$O in this example) using the corresponding mixing ratios, here assumed to be $\zeta_\text{H$_2$O} = 0.9$, $\zeta_\text{CO} = 0.1$, as shown in panel~b.}
\label{fig:ee_schematic}
\end{figure*}

We show a schematic illustration of this overlap method in \cref{fig:ee_schematic}, where the original $k$-distributions are shown in \cref{fig:rorr_schematic}a. From the $k$-coefficients of the minor absorber, in this case CO, an equivalent grey absorption is calculated according to \cref{eq:kBarApprox_net} or \cref{eq:kBarApprox_sw}, as illustrated in \cref{fig:ee_schematic}a. This grey absorption is then added onto the $k$-coefficients of the minor gas using the corresponding mixing ratios, here assumed to be $\zeta_\text{H$_2$O} = 0.9$ and $\zeta_\text{CO} = 0.1$, with the combined $k$-coefficients used in the radiation calculations shown in \cref{fig:ee_schematic}b.

\subsubsection{Determining the major absorber}

We consider two approaches for determining the major absorber in each band:
\begin{enumerate}
\item
Calculating the transmission of each gas down to the lower boundary using the maximum allowed mixing ratio for each gas given the elemental abundances, e.g. for H$_2$O assuming all O is in the form of H$_2$O. The gas with the smallest transmission in each band can be considered to be the major absorber. We label this approach EE, however, it is rather naive as local mixing ratios may significantly impact which gas dominates absorption in a band. Consequently, we have also considered a more sophisticated approach.
\item
In a given column the transmission of each gas is calculated from the top of the atmosphere down to the first layer where the total transmission is $< e^{-1}$, i.e. where the total optical depth has reached $1$ in the band. The major absorber is then defined as the gas with the smallest transmission at this level. We calculate the total transmission assuming random overlap and multiply individual gas transmissions according to \cref{eq:T_ab}. If the total optical depth does not reach $1$ we use the gas transmissions down to the lower boundary of the model instead. We label this approach AEE for adaptive equivalent extinction.
\end{enumerate}
To increase the efficiency and ease the implementation we use the average $k$-coefficients down to an optical depth of $1$ for each gas as defined in Eq.~(19) in \citet{Amundsen2014} in these calculations. Alternatively, the transmissions with AEE can be calculated using the local $k$-coefficients in each layer, however, we do not expect this to have a significant effect on the choice of major absorber.

\section{Application to hot Jupiter and brown dwarf atmospheres} \label{sec:application}

\begin{figure}
\centering
\includegraphics[width=\columnwidth]{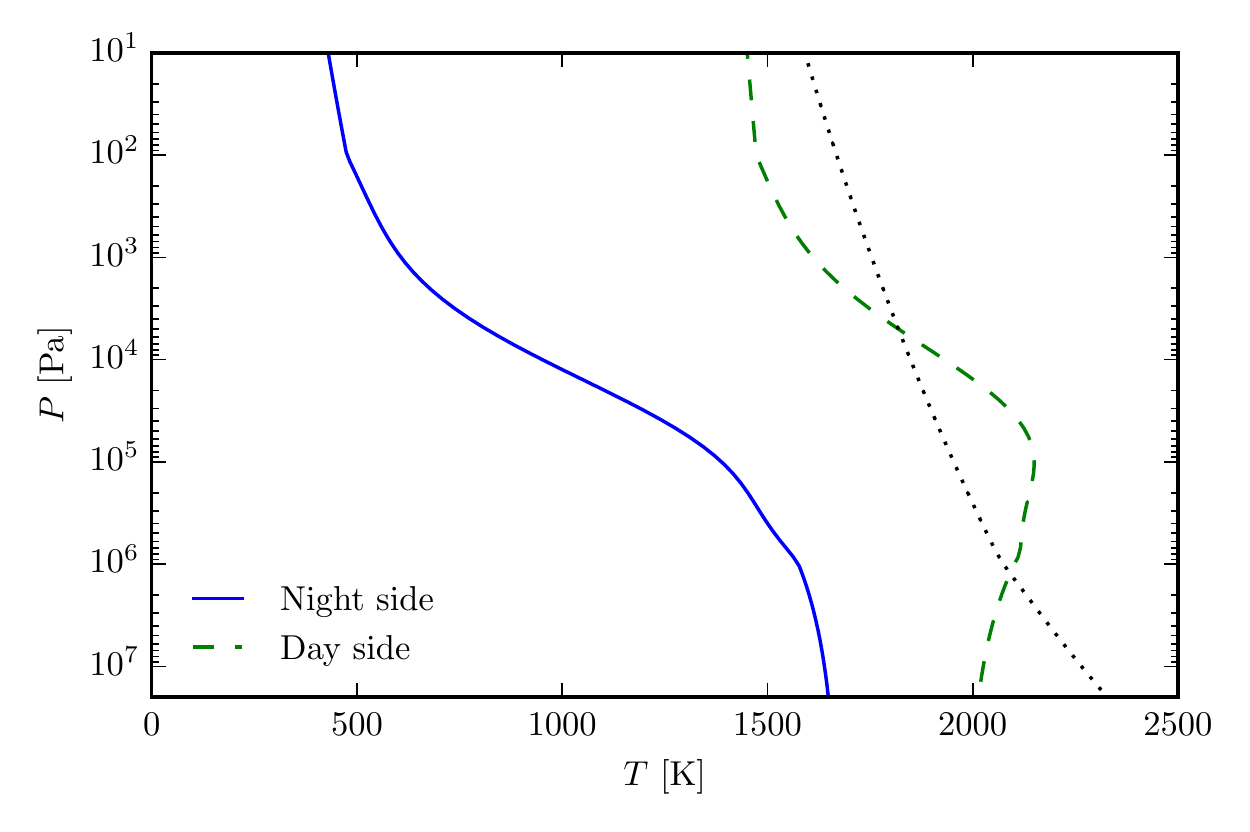}
\caption{$P$--$T$ profiles used for the tests in the present work. Profiles are derived from the equilibrium $P$--$T$ profile of \citet{Iro2005} as adopted by \citet{Cooper2005,Cooper2006}, \citet{Rauscher2010} and \citet{Heng2011} with the smoothing introduced by \citet{Mayne2014a}. We adopt the TiO/VO condensation curve from \citet{Fortney2008b}, plotted as a dotted line.}
\label{fig:pt_profiles}
\end{figure}

In this section we evaluate the accuracy of the overlapping gaseous absorption treatments discussed above when applied to hot Jupiter- and brown dwarf-like atmospheres by comparing them to line-by-line calculations. We have previously tested the applicability of the correlated-$k$ method to these atmospheres~\citep{Amundsen2014}, so we limit the discussion here to the overlap treatments. The test atmospheres here are identical to those in \citet{Amundsen2014}, however, in the present work we include all the updates to the opacities described in \citet{Amundsen2016a} and include $13$ opacity sources in total\footnote{We include the opacity of H$_2$O~\citep{Barber2006}, CO~\citep{Rothman2010}, CH$_4$~\citep{Yurchenko2014}, NH$_3$~\citep{Yurchenko2011}, H$_2$--H$_2$ and H$_2$--He collision induced absorption~\citep{Richard2012}, TiO~\citep{Plez1998}, VO~(B. Plez, priv. comm.), Li, Na, K, Rb and Cs~\citep{Heiter2008}.}. Abundances are as in \citet{Amundsen2016a}, and we include TiO and VO opacity for the day side $P$--$T$ profile, but not for the night side profile, in \cref{fig:pt_profiles}. We take the TiO/VO condensation curve from \citet{Fortney2008b} and apply a small smoothing of the abundance as described in \citet{Amundsen2016a}, with a smoothing scale of $\Delta T_\text{char}^i = \SI{10}{\kelvin}$.

An important aspect of the correlated-$k$ method is the choice of spectral bands. Both accuracy and computational costs increase with increasing number of bands, and it is therefore necessary to make a compromise between accuracy and speed. We adopt the $32$ spectral bands used in \citet{Amundsen2014} for all calculations using the correlated-$k$ method presented here, as this is enough for the error to become acceptably small, and small enough to facilitate use in both 1D and 3D models. A study of how the error varies depending on the number and placement of the spectral bands is beyond the scope of the present work.

The adopted $P$--$T$ profiles are shown in \cref{fig:pt_profiles}, which are derived from the equilibrium $P$--$T$ profile of \citet{Iro2005} derived for HD~209458b as adopted by \citet{Cooper2005,Cooper2006}, \citet{Rauscher2010} and \citet{Heng2011} with the minor adjustment introduced by \citet{Mayne2014a}. The TiO/VO condensation curve is plotted as dashed black line.

\subsection{Validity of the random overlap assumption} \label{sec:random_overlap}

In this section we test the validity of the random overlap assumption, which is considered to be more accurate than equivalent extinction~\citep{Edwards1996b}, by comparing it to line-by-line (LbL) calculations using our 1D atmosphere code \texttt{ATMO}~\citep{Amundsen2014,Tremblin2015,Tremblin2016}. Unfortunately the random overlap method without resorting and rebinning has not yet been implemented in \texttt{ATMO}. Instead we use $120$ rebinned $k$-terms in the tests presented in this section, which we have found to be more than sufficient for the solution to have converged (see \cref{sec:results} for convergence tests), in order to minimize errors caused by the rebinning. All line-by-line calculations were run at a resolution of $\SI{0.001}{\centi \metre^{-1}}$.

\subsubsection{Night side}

\begin{figure}
\centering
\includegraphics[width=\columnwidth]{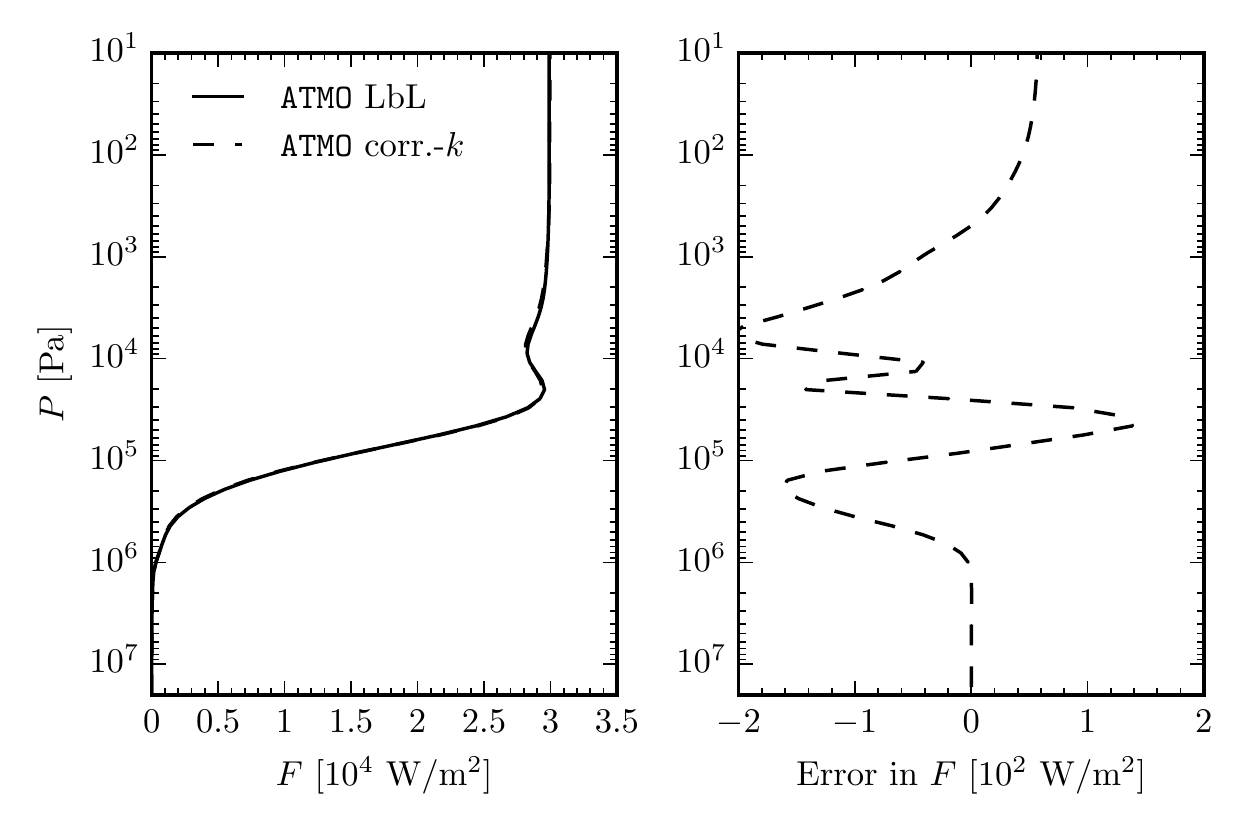}
\caption{Fluxes (left) and absolute errors in fluxes (right) for the night side $P$--$T$ profile in \cref{fig:pt_profiles}. The line-by-line (LbL) calculation was run at a $\SI{0.001}{\centi \metre^{-1}}$ resolution corresponding to $\num{5e7}$ monochromatic calculations, while we used $120$ rebinned $k$-coefficients in each band corresponding to \num{3840} pseudo-monochromatic calculations. Both results were obtained using \texttt{ATMO}.}
\label{fig:atmo_night_side_nflx}
\end{figure}

\begin{figure}
\centering
\includegraphics[width=\columnwidth]{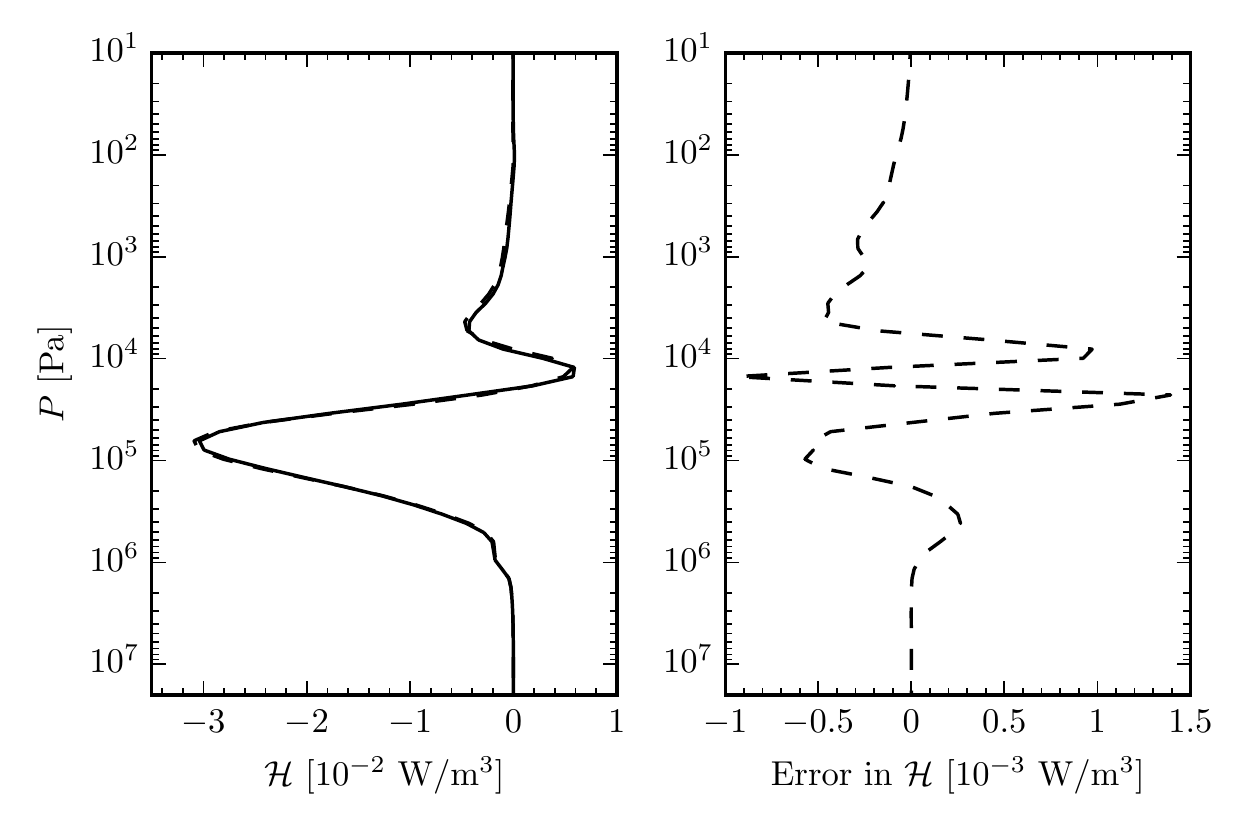}
\caption{Same as \cref{fig:atmo_night_side_nflx} but for heating rates.}
\label{fig:atmo_night_side_hrts}
\end{figure}

In \cref{fig:atmo_night_side_nflx,fig:atmo_night_side_hrts} we show the fluxes and heating rates, with corresponding errors calculated by comparing to line-by-line fluxes and heating rates, obtained for the night side $P$--$T$ profile shown in \cref{fig:pt_profiles}. It is clear that both fluxes and heating rates obtained when using the correlated-$k$ method with the random overlap method match the line-by-line results very well, with errors of a few percent. We note that these errors are both due to the use of the correlated-$k$ method and the random overlap assumption, and in agreement with the errors found in \citet{Amundsen2014}.

\subsubsection{Day side}

\begin{figure}
\centering
\includegraphics[width=\columnwidth]{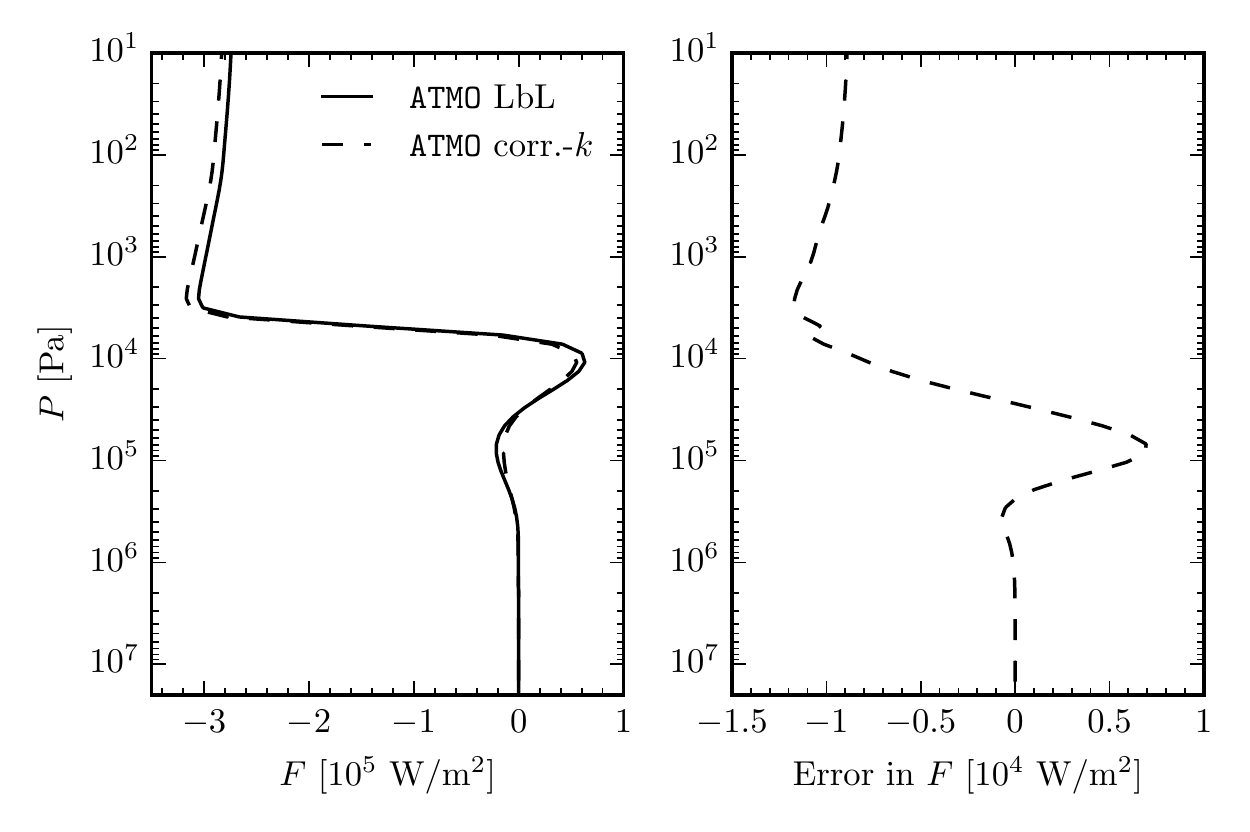}
\caption{Fluxes (left) and absolute errors in fluxes (right) for the day side $P$--$T$ profile in \cref{fig:pt_profiles}.}
\label{fig:atmo_day_side_nflx}
\end{figure}

\begin{figure}
\centering
\includegraphics[width=\columnwidth]{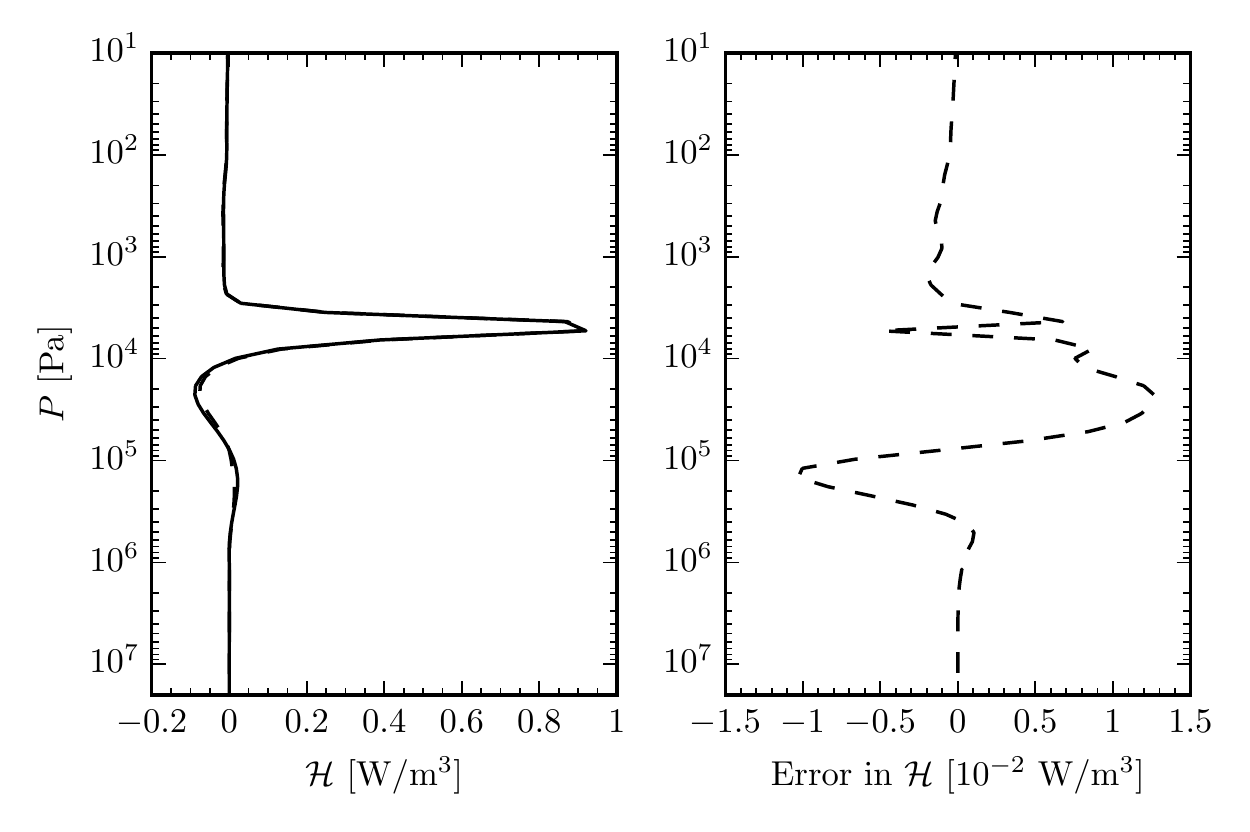}
\caption{Same as \cref{fig:atmo_day_side_nflx} but for heating rates.}
\label{fig:atmo_day_side_hrts}
\end{figure}

In \cref{fig:atmo_day_side_nflx,fig:atmo_day_side_hrts} we show the fluxes and heating rates, again with corresponding errors, for the day side $P$--$T$ profile shown in \cref{fig:pt_profiles}.  As for the night side $P$--$T$ profile both fluxes and heating rates obtained with the correlated-$k$ method using random overlap are in good agreement with the corresponding line-by-line results, with errors of a few percent. As for the night side these errors are both due to the use of the correlated-$k$ method and the random overlap assumption, and in agreement with the errors found in \citet{Amundsen2014}.

Based on these results we conclude that the random overlap method is indeed sufficiently accurate to be applied to these atmospheres for the bands adopted here. We note that for a different choice of bands, particularly for wider bands, the random overlap assumption should be reevaluated to make sure it is still valid.

\subsection{Comparison of gaseous overlap treatments} \label{sec:results}

Unfortunately, using $120$ rebinned $k$-terms becomes too computationally expensive even in our 1D atmosphere model. In this section we evaluate the accuracy of more efficient overlap treatments, and compare them in terms of both accuracy and computational efficiency using our GCM radiation scheme SOCRATES~\citep{Edwards1996a,Edwards1996b}. We have previously presented the adaptation of this radiation scheme to hot Jupiters~\citep{Amundsen2014}. The radiation scheme setup used here is identical to the one we use in our hot Jupiter GCM simulations presented in~\citet{Amundsen2016a}, i.e. for the two-stream approximation we use a diffusivity of $D = 1.66$ for thermal and $D = 2$ for stellar radiation. Rayleigh scattering is included for stellar radiation, and we ignore all other forms of scattering.

\subsubsection{Night side} \label{sec:night_side}

\begin{figure}
\centering
\includegraphics[width=\columnwidth]{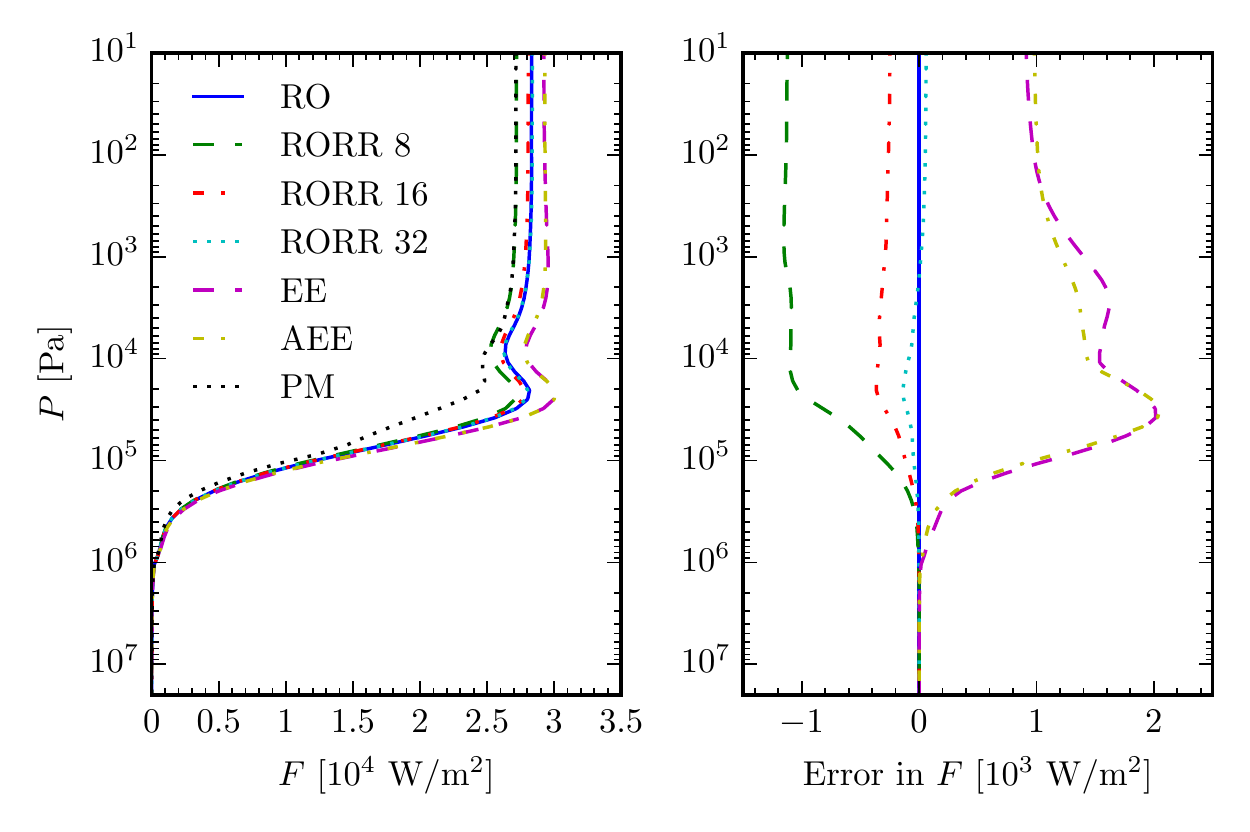}
\caption{Fluxes (left) and absolute errors in fluxes (right) obtained with the night side $P$--$T$ profile in \cref{fig:pt_profiles} using SOCRATES. Fluxes obtained using the random overlap method without resorting and rebinning (RO) are used to calculate errors for the random overlap with resorting and rebinning (RORR) with $n_k^\text{bin} = 8$, $16$ and $32$, equivalent extinction (EE), adaptive equivalent extinction (AEE) and pre-mixed opacities (PM).}
\label{fig:night_side_nflx}
\end{figure}

\begin{figure}
\centering
\includegraphics[width=\columnwidth]{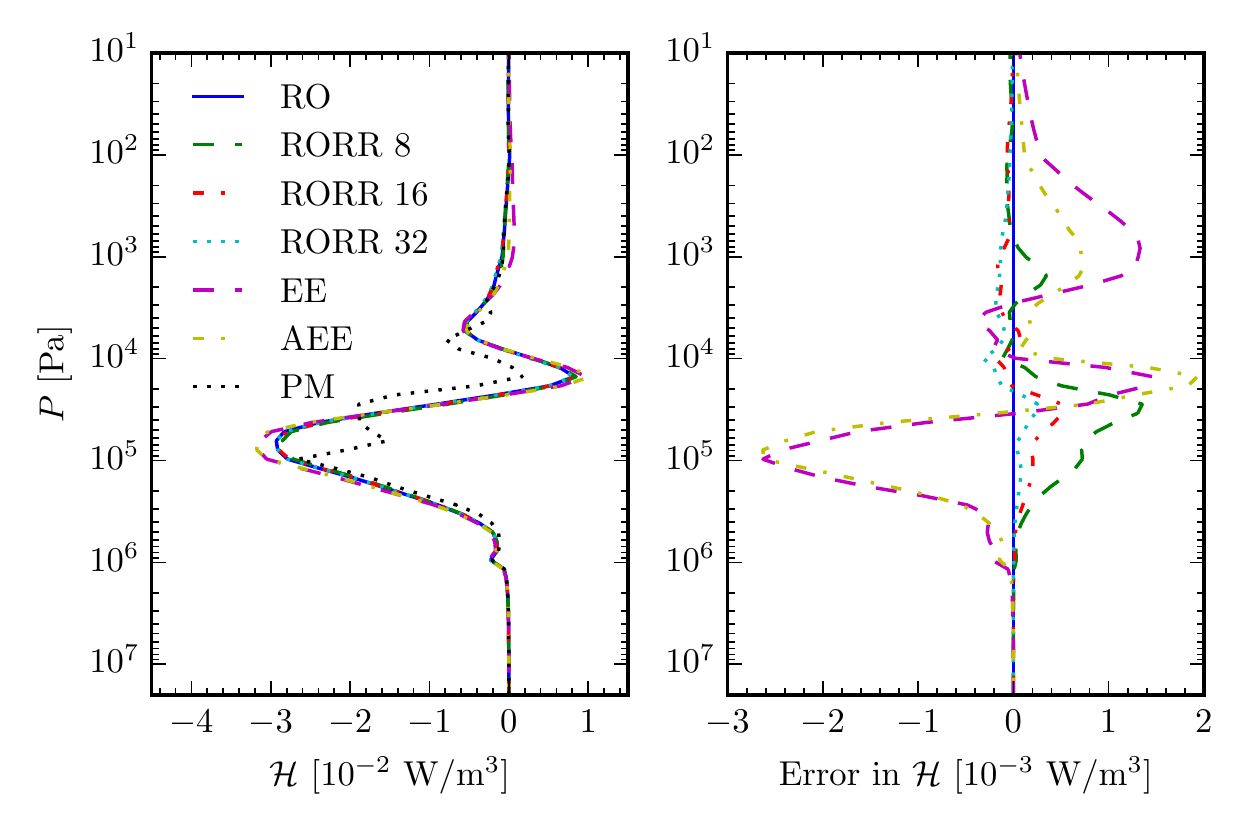}
\caption{Same as \cref{fig:night_side_nflx} but for heating rates. L1 norms of the errors \citep[see][the average heating rate error weighted by the local heating rates]{Amundsen2014} are \SI{4.5}{\percent} for RORR with $n_k^\text{bin} = 8$, \SI{1.9}{\percent} for RORR with $n_k^\text{bin} = 16$, \SI{1.5}{\percent} for RORR with $n_k^\text{bin} = 32$, \SI{13}{\percent} for EE, \SI{11}{\percent} for AEE and \SI{38}{\percent} for PM.}
\label{fig:night_side_hrts}
\end{figure}

We show in \cref{fig:night_side_nflx,fig:night_side_hrts} the thermal net upward fluxes and heating rates using the night side $P$--$T$ profile in \cref{fig:pt_profiles}, with corresponding errors, for all overlap treatments considered here. Errors are calculated by comparing to results obtained using the random overlap method without resorting and rebinning (RO). First, it is clear that using the random overlap method with resorting and rebinning (RORR) with an increasing number of rebinned $k$-terms $n_k^\text{bin}$ significantly decreases the error in both fluxes and heating rates. Equivalent extinction (EE) is somewhat less accurate than RORR with $n_k^\text{bin} = 8$, and with adaptive equivalent extinction (AEE) errors do not decrease significantly indicating the choice of major absorbers with EE was appropriate for this $P$--$T$ profile.

\begin{figure}
\centering
\includegraphics[width=\columnwidth]{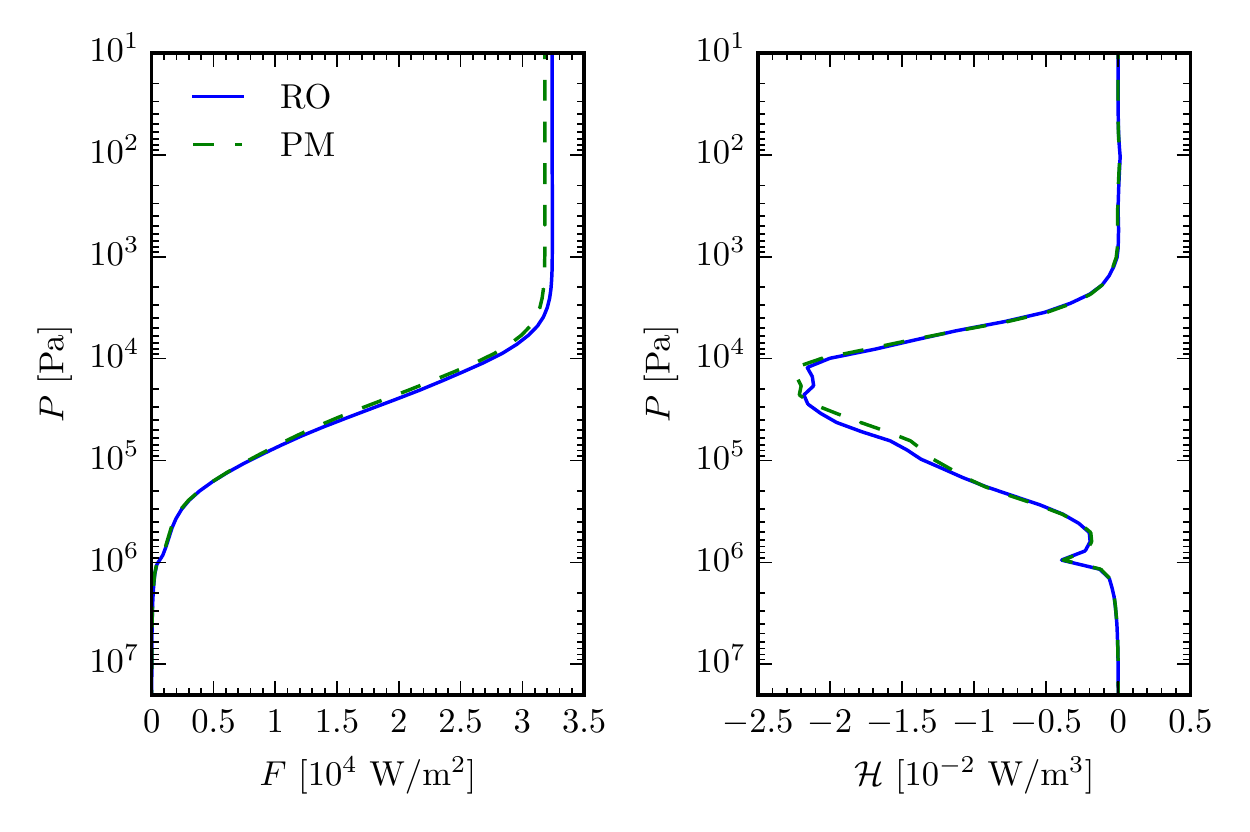}
\caption{Fluxes (left) and heating rates (right) obtained with the night side $P$--$T$ profile in \cref{fig:pt_profiles} using constant mixing ratios equal to the mixing ratios at $P = \SI{e4}{\pascal}$, $T = \SI{1000}{\kelvin}$. This eliminates errors caused by the implicit interpolation of mixing ratios with PM which dominates the errors seen using this overlap method in \cref{fig:night_side_nflx,fig:night_side_hrts}.}
\label{fig:night_side_const}
\end{figure}

Pre-mixed (PM) opacities are significantly less accurate than all the above overlap treatments, this stems from errors introduced by the interpolation in the pre-mixed opacity table. Changes in mixing ratios with temperature and pressure can cause large changes in the pre-mixed opacities which are not properly resolved by our opacity table. To illustrate this we have in \cref{fig:night_side_const} plotted fluxes and heating rates obtained again using the night side $P$--$T$ profile, but using constant mixing ratios equal to the mixing ratios at $P = \SI{e4}{\pascal}$, $T = \SI{1000}{\kelvin}$ both when computing the pre-mixed opacity table and when combining $k$-coefficients using RO. This eliminates errors caused by the implicit interpolation in mixing ratio with PM. The very small differences remaining between RO and PM are mainly due to small differences in the precision of the $k$-coefficients, which for RO are derived for each gas separately while for PM for the mixture directly. As in \citet{Amundsen2014} we use an opacity table logarithmically spaced in temperature and pressure, with $20$ temperature points between \SI{70}{\kelvin} and \SI{3000}{\kelvin} and $30$ pressure points between $\SI{e-1}{\pascal}$ and $\SI{e+8}{\pascal}$, with the opacity interpolation performed linearly in temperature. This is similar to the resolution used in previous works~\citep{Showman2009}.

In \cref{tbl:night_side} we give the relative computation times of the overlap treatments in \cref{fig:night_side_nflx,fig:night_side_hrts}. RO is, as expected, two to three orders of magnitude slower than the other overlap treatments. The quickest is PM, although (A)EE is only slightly slower. RORR, even with only 8 rebinned $k$-terms is about a factor of $3$ slower than (A)EE. We find that a significant fraction of the computation time with RORR is spent sorting the $k$-coefficients, and it is therefore important to use an efficient sorting algorithm. As mentioned in \cref{sec:rorr} we use a standard quicksort implementation, which we have found to consistently give good performance compared to shellsort and heapsort.

\begin{table}
\centering
\caption{Computation times of the thermal fluxes in SOCRATES for various overlap treatments using the night side $P$--$T$ profile in \cref{fig:pt_profiles} not including TiO and VO opacity, see discussion in \cref{sec:night_side}. The relative CPU computation time is the time relative to the fastest overlap method (PM).}
{\small
\begin{tabular}{l|r|r}
 & CPU time [\SI{e-2}{\second}] & Relative CPU time \\ \hline
RO & \num{1.1e3} & \num{1.7e3} \\
RORR 32 & \num{12.2} & \num{18.5} \\
RORR 16 & \num{5.0} & \num{7.6} \\
RORR 8 & \num{2.8} & \num{4.2} \\
(A)EE & \num{1.0} & \num{1.5} \\
PM & \num{0.66} & \num{1.0}
\end{tabular}
}
\label{tbl:night_side}
\end{table}


\subsubsection{Day side} \label{sec:day_side}

\begin{figure}
\centering
\includegraphics[width=\columnwidth]{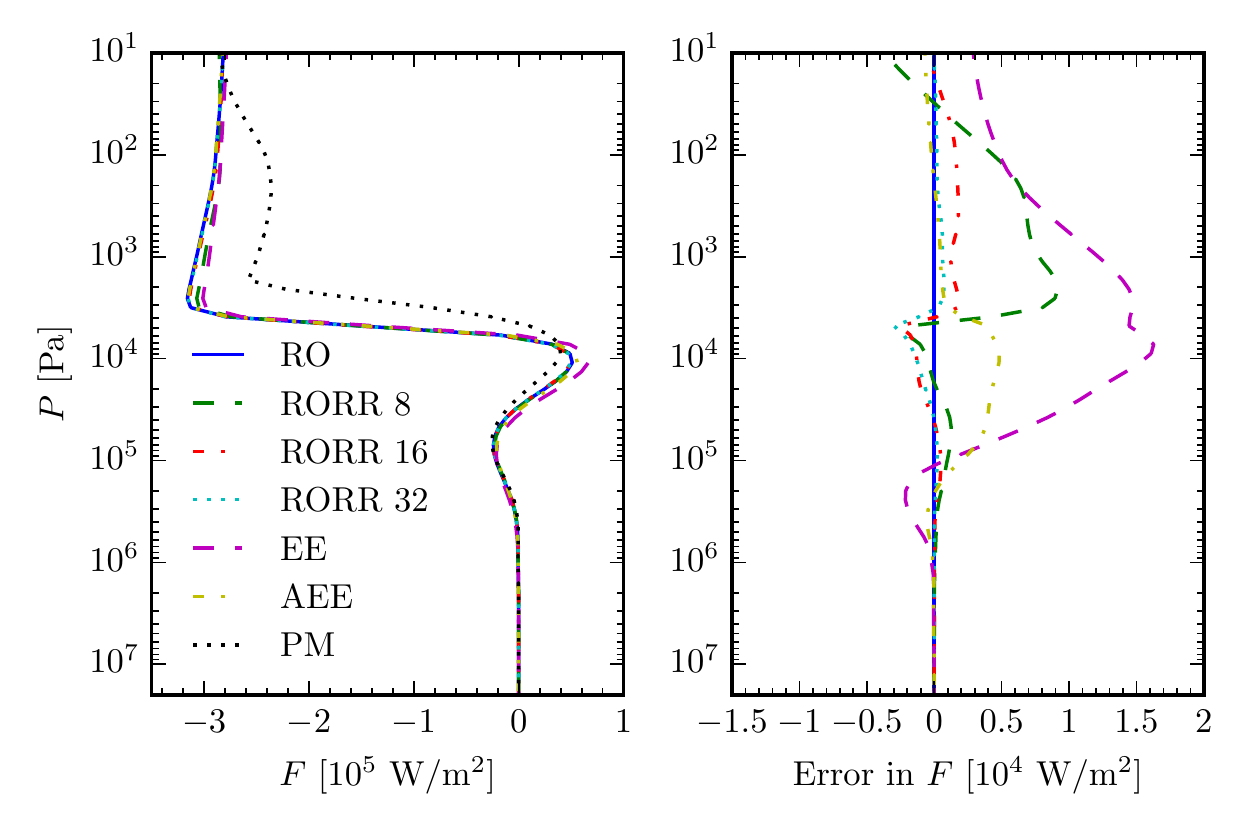}
\caption{Fluxes (left) and absolute errors in fluxes (right) obtained with the day side $P$--$T$ profile in \cref{fig:pt_profiles} using SOCRATES. Fluxes obtained using the random overlap method without resorting and rebinning (RO) are used to calculate errors for the random overlap with resorting and rebinning (RORR) with $n_k^\text{bin} = 8$, $16$ and $32$, equivalent extinction (EE), adaptive equivalent extinction (AEE) and pre-mixed opacities (PM).}
\label{fig:day_side_nflx}
\end{figure}

\begin{figure}
\centering
\includegraphics[width=\columnwidth]{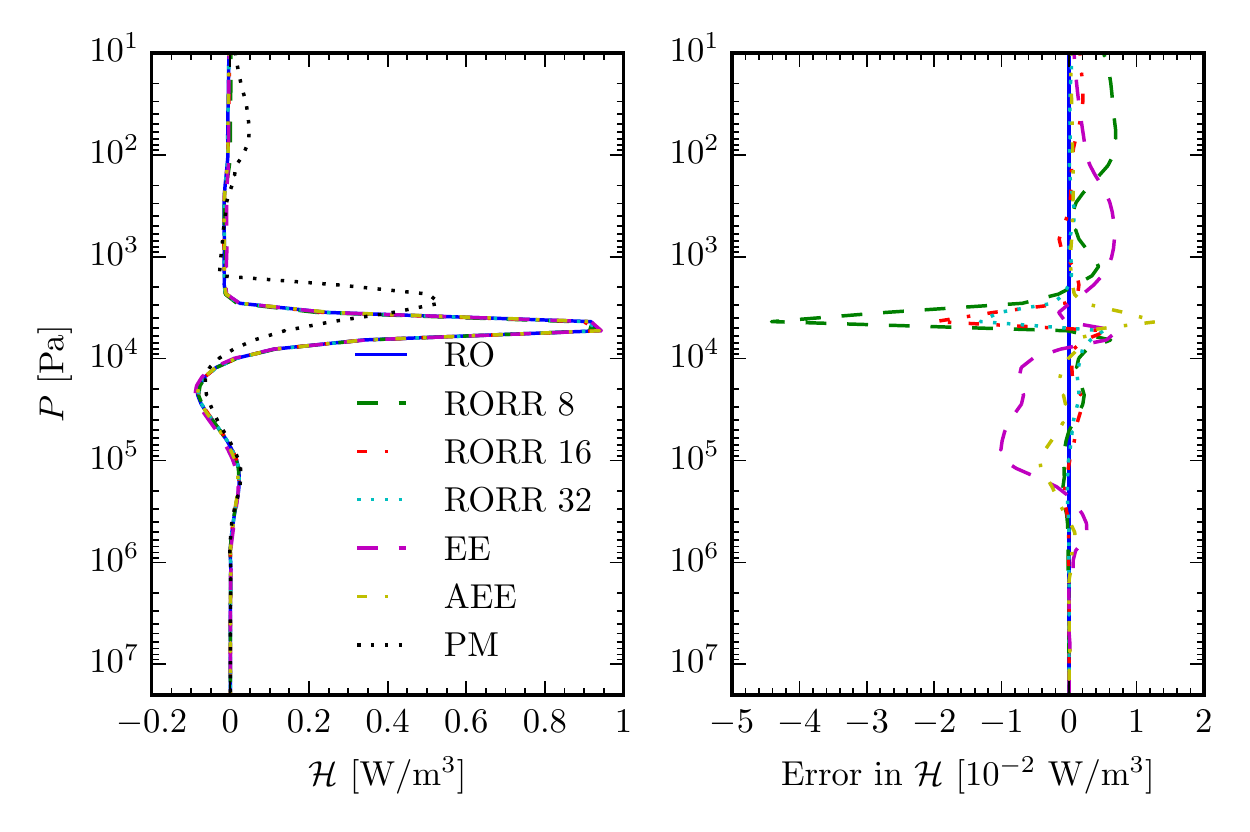}
\caption{Same as \cref{fig:day_side_nflx} but for heating rates. L1 norms of the errors (see the caption of \cref{fig:night_side_hrts}) are \SI{7.6}{\percent} for RORR with $n_k^\text{bin} = 8$, \SI{3.0}{\percent} for RORR with $n_k^\text{bin} = 16$, \SI{1.8}{\percent} for RORR with $n_k^\text{bin} = 32$, \SI{7.0}{\percent} for EE, \SI{2.2}{\percent} for AEE and \SI{119}{\percent} for PM.}
\label{fig:day_side_hrts}
\end{figure}

We show in \cref{fig:day_side_nflx,fig:day_side_hrts} total (thermal plus stellar) net upward fluxes and heating rates obtained using the day side $P$--$T$ profile in \cref{fig:pt_profiles}, with corresponding errors, for all overlap treatments considered here. Errors are, as for the night side, calculated by comparing to results obtained using RO. Results are overall similar to those obtained above for the night side, with errors being smallest for a large number of rebinned $k$-terms with RORR. A significant improvement in the accuracy is seen when using AEE compared to EE, indicating that the appropriate major absorbers have changed compared to the night side profile.

\begin{figure}
\centering
\includegraphics[width=\columnwidth]{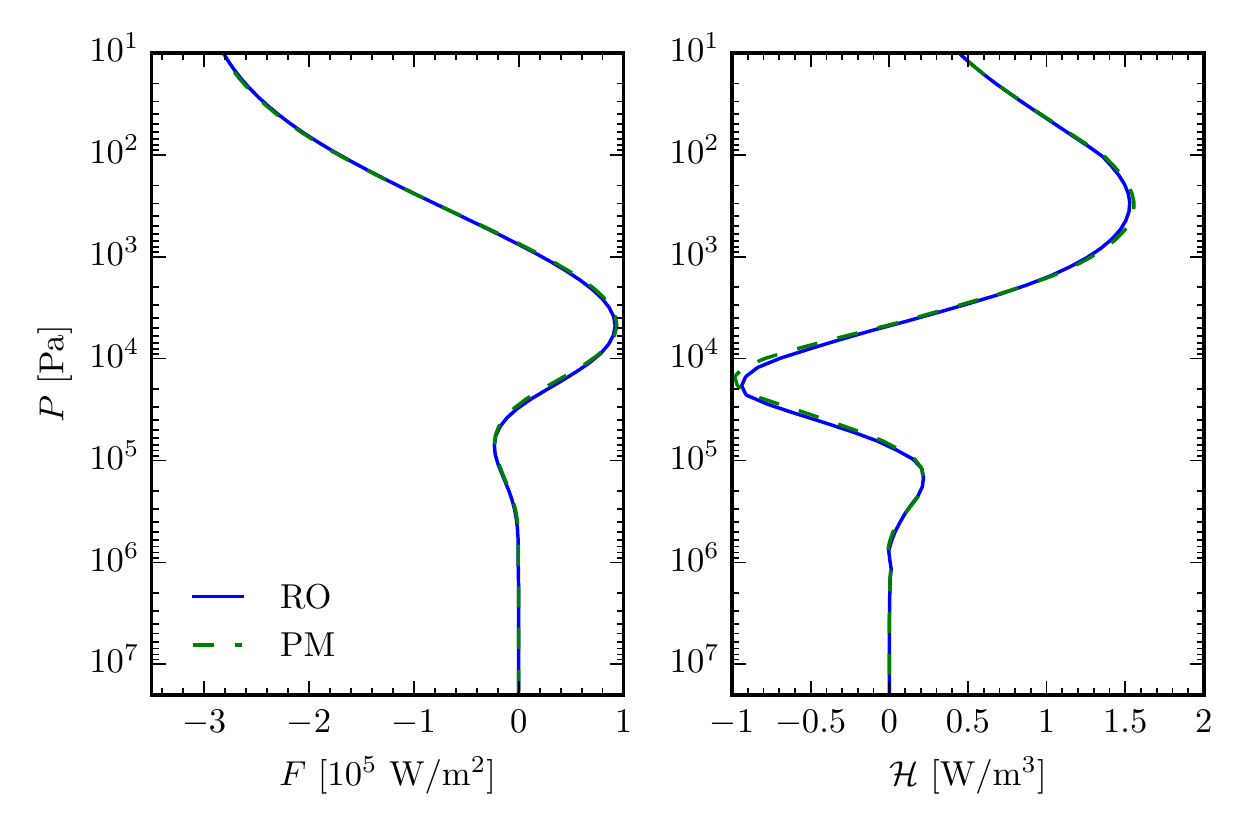}
\caption{Fluxes (left) and heating rates (right) obtained with the day side $P$--$T$ profile in \cref{fig:pt_profiles} using constant mixing ratios equal to the mixing ratios at $P = \SI{e4}{\pascal}$, $T = \SI{1900}{\kelvin}$. This eliminates errors caused by the implicit interpolation of mixing ratios with PM which dominates the errors seen using this overlap method in \cref{fig:day_side_nflx,fig:day_side_hrts}.}
\label{fig:day_side_const}
\end{figure}

Perhaps the most striking result is the large errors caused by using pre-mixed opacities, which are significantly larger for the day side compared to the night side. The flux changes very rapidly between \SI{e3}{\pascal} and \SI{e4}{\pascal}, which causes a large increase in the heating rate. Looking at \cref{fig:pt_profiles} this discontinuity occurs as the $P$-$T$--profile crosses the condensation curve of TiO and VO. Both molecules are strong absorbers in the visible, and the presence of these molecules leads to a strong absorption of the incoming stellar radiation. The steep vertical gradient in the mixing ratios of TiO and VO when the temperature is near the condensation temperature causes a similarly steep gradient in the opacity. When using PM this transition is smoothed out as the resolving power is limited by the number of $P$--$T$ points in the look-up $k$-coefficient table, thus reducing the accuracy of the interpolation. To illustrate this we have in \cref{fig:day_side_const} plotted fluxes and heating rates obtained again using the day side $P$--$T$ profile, but with constant mixing ratios equal to the mixing ratios at $P = \SI{e4}{\pascal}$, $T = \SI{1900}{\kelvin}$, similar to \cref{fig:night_side_const} for the night side, thereby eliminating errors caused by the implicit interpolation in mixing ratio with PM.

In \cref{tbl:day_side_lw,tbl:day_side_sw} we give the computation times for the different overlap treatments. Results are similar to those obtained for the night side in \cref{tbl:night_side} with PM being the fastest, (A)EE slightly slower, and RORR with only $8$ rebinned $k$-terms about $3$ times slower than (A)EE.

\begin{table}
\centering
\caption{Computation times of the thermal fluxes in SOCRATES for various overlap treatments using the day side $P$--$T$ profile in \cref{fig:pt_profiles} including TiO and VO opacity, see discussion in \cref{sec:day_side}. The relative CPU time is the computation time relative to the fastest overlap method (PM).}
\begin{tabular}{l|r|r}
 & CPU time [\SI{e-2}{\second}] & Relative CPU time \\ \hline
RO & \num{4.2e4} & \num{5.8e4} \\
RORR 32 & \num{14.9} & \num{20} \\
RORR 16 & \num{6.3} & \num{8.6}  \\
RORR 8 & \num{3.5} & \num{3.8} \\
(A)EE & \num{1.3} & \num{1.8} \\
PM & \num{0.73} & \num{1.0}
\end{tabular}
\label{tbl:day_side_lw}
\end{table}


\begin{table}
\centering
\caption{Computation times of the stellar fluxes in SOCRATES for various overlap treatments using the day side $P$--$T$ profile in \cref{fig:pt_profiles} including TiO and VO opacity, see discussion in \cref{sec:day_side}. The relative CPU time is the computation time relative to the fastest overlap method (PM).}
\begin{tabular}{l|r|r}
 & CPU time [\SI{e-2}{\second}] & Relative CPU time \\ \hline
RO & \num{5.5e4} & \num{9.2e4} \\
RORR 32 & \num{14.6} & \num{24} \\
RORR 16 & \num{5.8} & \num{9.7} \\
RORR 8 & \num{3.0} & \num{5.0} \\
(A)EE & \num{1.0} & \num{1.7} \\
PM & \num{0.60} & \num{1.0}
\end{tabular}
\label{tbl:day_side_sw}
\end{table}


\section{Conclusions} \label{sec:conclusions}

We have evaluated the applicability of several gaseous overlap treatments in hot Jupiter atmosphere models. We have shown that the random overlap method gives good accuracy and flexibility, but without resorting and rebinning (RO) it is too slow for practical use. With resorting and rebinning (RORR) it becomes much more computationally efficient, and the accuracy and speed can be adjusted by varying the number of rebinned $k$-terms.

Equivalent extinction (EE) is about three times faster than RORR with only $8$ rebinned $k$-terms, and benefits from the same flexibility as RORR, however, it is clear that particular care must be taken when choosing the major absorber in each band. We present one way of determining the major absorber, which we term adaptive equivalent extinction (AEE), that benefits from better accuracy compared to a more naive choice without a major loss of computational efficiency.

The fastest overlap treatment considered here is pre-mixed opacities, however, it lacks the flexibility of the random overlap method and equivalent extinction. In addition, particular care must be taken if there are regions of the atmosphere where the total opacity changes rapidly as a function of height to ensure steep variations in composition at important chemical equilibrium boundaries are adequately resolved. If these are not adequately resolved, as is the case in our pre-mixed table, interpolation can cause such transitions to be significantly smoothed out. We have shown that TiO in particular can be a significant cause of inaccuracies, but other species such as H$_2$O, CH$_4$ and CO may also lead to inaccuracies as seen in \cref{sec:night_side}. A lower resolution can be tolerated in $k$-coefficient tables of individual gases than in pre-mixed tables as individual opacities vary more smoothly.

In our 1D atmosphere code \texttt{ATMO} we use RORR, usually with about 30 rebinned $k$-terms in each band. This gives us the flexibility to manipulate gas mixing ratios without recomputing or having to add additional dimensions to the $k$-table. It also uniquely allows us to treat non-equilibrium chemistry consistently, i.e. have non-equilibrium abundances feed back on the total opacity and consequently the calculated $P$--$T$ profiles~\citep{Tremblin2015,Tremblin2016,Drummond2016}.

Unfortunately, RORR, even with only 8 rebinned k-coefficients, is too slow for use in our GCM coupling dynamics and radiative transfer, and we consequently use equivalent extinction. The method for adaptively choosing the major absorber in each band (AEE) has not yet been implemented in the radiation scheme coupled to our GCM. In \citet{Amundsen2016a} we therefore use the simple approach of determining the major absorber in each band instead (EE). Work is in progress to improve the treatment of overlapping gaseous absorption in our GCM, one possibility being using EE for bands with a clear major absorber and RORR for bands with more than one significant absorber. In addition, as the current definition of the equivalent extinction for the stellar component relies on direct stellar fluxes at the bottom boundary it will become important to improve this definition of the equivalent extinction when the diffuse stellar flux becomes more important, e.g. when including stronger short-wave scattering.

\begin{acknowledgements}
We thank the referee, Mark Marley, for comments that significantly improved the paper. This work is partly supported by the European Research Council under the European Community's Seventh Framework Programme (FP7/2007-2013 Grant Agreement No. 247060-PEPS and grant No. 320478-TOFU). D.S.A. acknowledges support from the NASA Astrobiology Program through the Nexus for Exoplanet System Science. J.M. acknowledges the support of a Met Office Academic Partnership secondment. The calculations for this paper were performed on the DiRAC Complexity machine, jointly funded by STFC and the Large Facilities Capital Fund of BIS, and the University of Exeter Super-computer, a DiRAC Facility jointly funded by STFC, the Large Facilities Capital Fund of BIS and the University of Exeter.
\end{acknowledgements}

\appendix

\section{Relationship between convolution and random overlap} \label{sec:convolution}

The transmission through a homogeneous slab is, from \cref{eq:transmission_def},
\begin{align}
\mathcal T(u) &= \int_{\nut_1}^{\nut_2} \md \nut \, w(\nut) e^{-k(\nut) u}
= \int_0^\infty \md k \, f(k) e^{-k u} \\
&= \int_0^1 \md g \, e^{-k(g) u},
\end{align}
where $f(k) \, \md k$ is the probability of the absorption coefficient being in the interval $[k, k + \md k]$ taking into account the weighting $w(\nut)$, and $g(k)$ is the cumulative probability distribution
\begin{equation}
g(k) = \int_0^k \md k' \, f(k') .
\end{equation}
For each absorbing gas $i$, the opacity probability distribution $f_i(k_i)$ can be derived, and need to be combined using the respective mixing ratios $\zeta_i$. An exact procedure for doing this does not exist as the wavelength information is lost when deriving $f_i(k_i)$. Assuming that random variables picked using these probability distributions are independent, however, they can be convolved to get the combined probability distribution $f(k)$. We restrict the discussion here to combining the opacity distributions of two gases, as the procedure can easily be repeated to include an arbitrary number of gases.

In order to perform the convolution we need to take into account the mixing ratios of the gases, i.e. we need to find $f_i'(k_i')$ where $k_i' = \zeta_i k_i$. We know that
\begin{equation}
f_i'(k_i') \, \md k_i' = f_i(k_i) \, \md k_i,
\end{equation}
which yields
\begin{equation}
f_i'(k_i') = f_i(k_i) \frac{\md k_i}{\md k_i'} = \frac{f_i(k_i'/\zeta_i)}{\zeta_i}.
\end{equation}

Having derived the opacity distributions $f_x(k_x)$ and $f_y(k_y)$ of two different gases $x$ and $y$ with mixing ratios $\zeta_x$ and $\zeta_y$, the total opacity distribution will be given by the convolution of the two taking their respective mixing ratios into account:
\begin{align}
f_{xy}'(k_{xy}') &= \int_0^\infty \md k_y' f_x'(k_{xy}' - k_y') f_y'(k_{y}') \\
&= \frac{1}{\zeta_x \zeta_y} \int_0^\infty \md k_y' f_x([k_{xy}' - k_y']/\zeta_x) f_y(k_{y}'/\zeta_y).
\end{align}
The total transmission becomes
\begin{align}
\mathcal T(u) &= \int_0^\infty \md k_{xy}' \, f_{xy}'(k_{xy}') e^{k_{xy}' u} \\
&= \frac{1}{\zeta_x \zeta_y} \int_0^\infty \md k_{xy}' \, e^{k_{xy}' u} \int_0^\infty \md k_y' f_x([k_{xy}' - k_y']/\zeta_x) f_y(k_y'/\zeta_y) .
\end{align}
Changing the integration variable of the second integral to $k_y = k_y'/\zeta_y$ yields
\begin{equation}
\mathcal T(u) =
\frac{1}{\zeta_x} \int_0^\infty \md k_{xy}' \, e^{k_{xy}' u} \int_0^\infty \md k_y f_x([k_{xy}' - k_y']/\zeta_x) f_y(k_y). 
\end{equation}
Similarly changing the integration variable of the first integral to $k_x = k_x'/\zeta_x = [k_{xy}' - k_y']/\zeta_x$, and using\begin{equation}
k_{xy}' = k_x' + k_y' = k_x \zeta_x + k_y \zeta_y,
\end{equation}
yields
\begin{align}
\mathcal T(u) &=
\int_0^\infty \md k_x \int_0^\infty \md k_y f_x(k_x) f_y(k_y) e^{(k_x \zeta_x + k_y \zeta_y) u} \\
&= \int_0^\infty \md k_x f_x(k_x) e^{k_x \zeta_x u} \times \int_0^\infty \md k_y f_y(k_y) e^{k_y \zeta_y u} \\
&= \mathcal T_x (u_x) \times \mathcal T_y (u_y).
\label{eq:T_xy_convolution}
\end{align}
\Cref{eq:T_xy_convolution} is identical to \cref{eq:T_ab}, i.e. the random overlap method is equivalent to convolving the opacity probability distributions.

\bibliographystyle{aa}
\bibliography{/Users/damundse/Documents/bibliography_hot_Jupiters}

\end{document}